\documentclass[aps,prc,showpacs,showkeys,superscriptaddress,nofootinbib,twocolumn,floatfix]{revtex4}
\usepackage{bm}
\usepackage{placeins}
\usepackage{float}
\usepackage{amsfonts}
\usepackage{multirow}
\usepackage{graphicx,color,amsmath,amssymb}
\usepackage[version=4]{mhchem}
\usepackage{hyperref}
\usepackage{color,soul}

\newcommand{\eg}{\textit{e.g.}}

\newcommand{\ie}{\textit{i.e.}}

\begin{document}

\title{Thermalization of Nuclear Matter in Heavy-Ion Collisions at Fermi Energies}

\author{Thomas Onyango}
\affiliation{Cyclotron Institute and Department of Physics and Astronomy, Texas A$\&$M University, College Station, TX 77843-3366, USA}
\author{Aldo Bonasera}
\affiliation{Cyclotron Institute and Department of Physics and Astronomy, Texas A$\&$M University, College Station, TX 77843-3366, USA}
\affiliation{Laboratori Nazionali del Sud, INFN, via Santa Sofia, 62, 95123 Catania, Italy}
\author{Ralf Rapp}
\affiliation{Cyclotron Institute and Department of Physics and Astronomy, Texas A$\&$M University, College Station, TX 77843-3366, USA}

\date{\today}

\begin{abstract}
We analyze the time evolution of the kinetic properties of nuclear matter produced in heavy-ion collisions at Fermi energies. The collision system is 
simulated using Constrained Molecular Dynamics (CoMD) transport calculations whose output is the isospin, position, and momentum of the nucleons. 
Focusing on central 35\,A$\cdot$MeV $^{40}$Ca+$^{40}$Ca collisions we utilize this information to extract localized momentum distributions in volume 
elements of 8\,fm$^3$ and time steps of 5\,fm/$c$. We then parameterize the single-particle momentum distributions with thermally motivated fit 
functions in the local rest frame of each cell. While the transverse-momentum distributions are well reproduced by thermal ones, the longitudinal 
ones carry a marked imprint of the initial nuclear motion which we capture by introducing a centroid motion into our fit functions. In particular, we find 
that Fermi distributions yield significantly better fits than Boltzmann ones, a consequence of the Pauli blocking implemented in CoMD. From the fits we 
extract the time dependence of the thermodynamic and collective properties of the excited nuclear medium. 
We find that the transverse temperature gradually rises to about 6\,MeV, which is accompanied by a dissipation of the initial centroid motion of the incoming nuclei which vanishes at about 100\,fm/$c$ after initial impact. We are therefore able to track the transition of beam energy into random 
kinetic energy for nucleons, suggesting a three-dimensional equilibration of energy in the late stages of the collision.
\end{abstract}
\maketitle

\section{Introduction}
\label{sec_intro} 
The study of heavy-ion collisions (HICs) aims at obtaining systematic insights into the behavior of nuclear matter under various conditions of density and temperature. Analyses of such collisions can yield evidence for a local thermalization of the medium and thus enable the study of the phase diagram of strongly interacting matter, in particular its transport and microscopic properties and the occurrence of possible phase transitions. Heavy-ion collisions have been carried out extensively at ultrarelativistic collision energies where the system appears to reach local thermal equilibrium~\cite{Stock:2010hoa}. The situation is less clear at lower bombarding energies, $E_\text{Lab}<1$\,GeV/A~\cite{COLONNA2020103775}. Widely used methods to extract temperatures in HICs include the analysis of momentum spectra~\cite{Siemens:1978pb,Danielewicz:1998vz,Heinz:2013th}, chemical compositions~\cite{Braun-Munzinger:2003pwq,Ropke:2013hha}, fluctuation observables~\cite{Stephanov:1999zu,Friman:2011pf,Zheng:2014afa} and electromagnetic radiation which can, in principle, penetrate out from deeper in the medium~\cite{Gale:2009gc,Rapp:2014hha}. Under conditions where thermalization is questionable, coarse-graining methods have been applied to microscopic transport calculations, by discretizing the medium evolution into finite spatial cells and time steps, see, \eg, Refs.~\cite{Huovinen:2002im,Santini:2011zw,Endres:2014zua,Galatyuk:2015pkq}. This allows to test whether the local distributions reach near local thermal equilibrium and extract local thermodynamic parameters that can subsequently be used, \eg, to compute emission spectra of electromagnetic radiation. A more global variant of this method is to average over larger regions of the nuclear fireball to determine the nucleon number and energy density, and then map these quantities into a temperature and chemical potential~\cite{Zhou:2012bd,Deng:2016pqu}. 

In the present work we focus on a coarse-graining analysis of transport simulations of HICs at Fermi energies. At these energies, the nucleon chemical potential is typically much larger than the temperature,  $\mu_N\gg T$, and thus Pauli blocking is expected to be pertinent. There are various transport approaches to simulate HICs at these energies that can essentially be placed into two groups, \ie, semiclassical ones based on the Boltzmann equation (Boltzmann-Uehling-Uhlenbeck (BUU) or Boltzmann-Vlasov models) and Quantum Molecular Dynamics (QMD), see 
Refs.~\cite{Bertsch:1988ik,Aichelin:1989pc,Khoa:1992vhc,Bonasera:1994zz,Zhang:2017esm} for reviews and comparisons. The BUU models typically utilize in-medium nucleon-nucleon cross sections together with effective mean fields to determine the evolution of the nucleon distributions. They do not explicitly implement Pauli blocking, but large repulsive mean fields at short distances effectively avoid nucleons coming close to each other in phase space. On the other hand, in QMD models the system evolves based on two-body interactions and fluctuating nucleons in phase space treated via Gaussian wave packets. Anti-symmetrized Molecular Dynamics (AMD) has an extra feature in that it has a mechanism for anti-symmetrizing the wavefunctions of the nucleons in coordinate space~\cite{Horiuchi:1994hnt}. Furthermore, Fermionic Molecular Dynamics (FMD) utilizes Fermi-Dirac statistics on the many-body level~\cite{Feldmeier:1989st} and also has a strong repulsion when nucleons are close in phase space. There is also an update to QMD called Isospin-dependent Quantum Molecular Dynamics (IQMD) which improves upon QMD by taking into account the spin of the nucleons (and including Pauli blocking)~\cite{Zhou:2012bd}. Here we employ a version of QMD transport that implements Pauli blocking yet supplies microscopic information about the nucleons throughout the collision, referred to as Constrained Molecular Dynamics (CoMD)~\cite{Papa:2000ef,Zheng:2014afa}. In particular, CoMD achieves a fair description of the ground-state properties of nuclei which is relevant for the initial state of the collision. 

The paper is organized as follows. In Sec.~\ref{sec_method}, we briefly recall the basic features of CoMD and set up our procedure for coarse graining to 
extract local nucleon momentum distributions. In Sec.~\ref{sec_extract}, we describe the process for fitting these distributions using thermally motivated 
fit functions which, however, account for off-equilibrium effects in the incoming-beam direction which are essential to account
for the remnants of the initial beam momentum. In Sec.~\ref{sec_evo}, we extract the thermodynamic parameters from the fit functions including 
the time dependence of the off-equilibrium effects which allows us to interpret the dissipation and transferal of the incoming beam energy into thermal motion. 
In Sec.\ref{sec_concl}, we conclude and give an outlook for future applications of our approach.

\section{Constrained Molecular Dynamics Model and Coarse Graining}
\label{sec_method} 
Constrained Molecular Dynamics is a transport model designed to simulate HICs at low and intermediate energies. The nucleons are not treated as point-like particles but as Gaussian wave packets with a specified width in phase space during the evolution of the collision system. Pauli blocking is also implemented as a constraint. As CoMD is a microscopic model, pre-equilibrium effects are automatically included if allowed by the dynamics. Due to its $N$-body nature as a QMD-type transport model, CoMD can account for multi-fragmentation, fission, and evaporation at different time scales. In particular, fragmentation may occur at late times, $\sim$300-400\,fm/$c$, but it is rather sensitive to the method used to recognize the fragments. The version of CoMD used in the present work utilizes the minimum spanning method where nucleons belong to a fragment if their relative distance is smaller than the width of the gaussian nucleon distribution in coordinate space. In order to assess the impact of pre-equilibrium emission, we could in principle vary the equation of state (EoS) through the nucleon-nucleon cross section, which is, however, beyond the scope of the present paper.

\begin{figure}[t!]
\begin{minipage}{\linewidth}
\centering
\includegraphics[width=\linewidth]{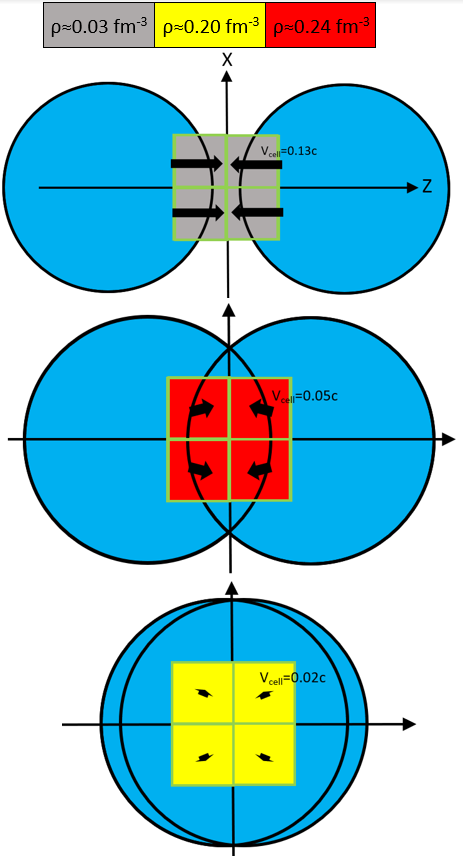} 
\end{minipage}
\caption{Schematic diagram showing the location of the 8 cells adjacent to the center of a central Ca-Ca(35\,A$\cdot$MeV) collision used for the coarse-graining analysis for different time snapshots: just before the nuclei start colliding, at $t\approx40$\,fm/$c$ (upper panel); 
shortly after the nuclear collisions have commenced, at $t\approx55$\,fm/$c$ (middle panel); and close to the full geometric overlap of the colliding nuclei at $t\approx 70$\,fm/$c$ (lower panel). The color of the cell denotes the calculated density at that time corresponding to the values given in the color scale at the top of the figure. The arrows show the direction and magnitude of the average velocity of the nucleons within each cell.} 
\label{fig_schematic}
\end{figure}

Originally designed to study mass fragments that have evaporated from the final state of nuclear collisions, we have modified the code to
output at each timestep the spin, isospin, coordinates, and momenta of the centers of individual nucleons instead of clusters of nucleons. 
We have generated 24,000 CoMD events to simulate central collisions of \ce{^{40}Ca}+\ce{^{40}Ca} at 35\,A$\cdot$MeV beam energy. 
The simulations are initialized by randomly rotating the ground state of the projectile and target in coordinate and momentum space. 
Once we have the output of the transport code, we perform a coarse graining by discretizing space into cells of volume (2\,fm)$^3$ and time into intervals of 5\,fm/$c$. This allows us to create localized momentum spectra. Instead of a single central cell we use data from each cell in the octant adjacent to the  center of the collision (8 cells total), see Fig.~\ref{fig_schematic}.

\begin{figure}[t!]
\begin{minipage}{1.0\linewidth}
\centering
\includegraphics[width=\linewidth]{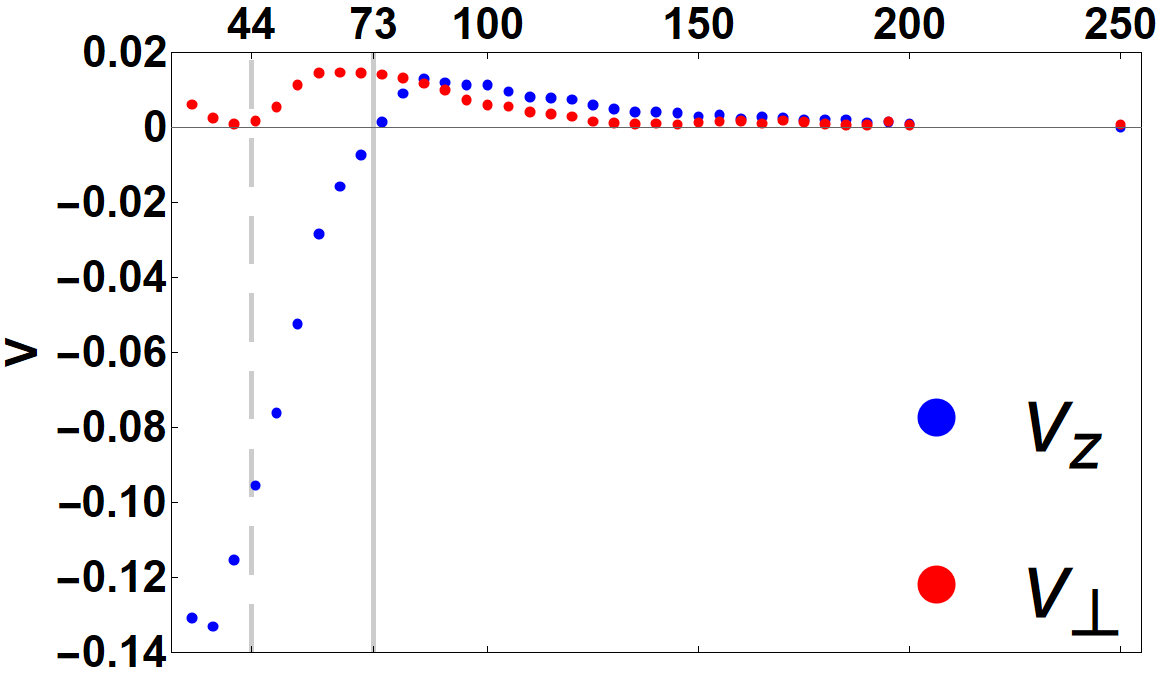}
\end{minipage}
\caption{Time evolution of the longitudinal velocity component and transverse speed of the cells from the positive-$z$ half-space. The velocities of the cells from negative-$z$ space have the same behavior except that the $v_z$ component has a different sign (and the transverse components are oriented away from the origin).}
\label{fig_velocity}
\end{figure}

The momentum distribution from each cell (averaged over all events) at each timestep is corrected for the center-of-mass (CM) motion of the cells to ensure 
the extraction of the thermodynamic quantities in their local rest frames. This is done by calculating the average velocity of all nucleons in the cell and then
subtracting that from the individual velocities of the nucleons in the cell. From Fig.~\ref{fig_schematic} one recognizes a symmetry where the $z$-components of the cells' velocities are equal-opposite to each other in the positive vs.~negative $z$-regions of space, while the transverse speeds, $v_\perp$, are identical in all cells (modulo statistical fluctuations) with their $x$- and $y$-components carrying the sign of the corresponding spatial coordinate axis (\ie, directed away from the origin).  
The time evolution of the average longitudinal and transverse velocity components of cells from the positive-$z$ half-space is illustrated in 
Fig.~\ref{fig_velocity}. 

The momentum distributions are fitted with a parameterized function as described in the following section, allowing the extraction of a chemical potential and temperature, which ultimately gives access to the time evolution of the thermodynamic properties of the nuclear medium in the center of the collision (note that the cell size is much smaller than the diameter of the $^{40}$Ca nucleus). 

Our method for studying the time evolution of nuclear collisions differs from other methods thus far employed in the literature. Specifically, quadrupole fluctuations in transverse momentum have been used in Refs. \cite{Zheng:2013qhq,Mabiala:2014hxa,Deng:2016pqu,Liu:2017vlx}, with a variance defined as
\begin{equation}
<\sigma_{xy}^{\ 2}>\equiv\frac{\int{d^3p(p_x^2-p_y^2)^2f(p)}}{\int{d^3pf(p)}} \ .
\label{eq_quadrupole}
\end{equation}
From the variance and number density, the temperature and chemical potential have been calculated. This method does not explicitly address phenomena occurring in the longitudinal direction due to the initial motion of the incoming nuclei.
Another method is the hot Thomas-Fermi formalism~\cite{Khoa:1992vhc}. In Ref.~\cite{Zhou:2012bd}, based on simulations with IQMD, a single central sphere of radius 5\,fm in the center-of-mass frame has been employed to analyze \ce{Au}+\ce{Au} collisions using suitably extracted bulk properties of the medium (number density, kinetic-energy density and entropy density). The analyzed volume is larger than in our case, and no explicit fits to the nucleon distributions were carried out.

\section{Extraction of Distribution Functions}
\label{sec_extract}
We carry out our fits to 3-dimensional distribution functions in a two-step procedure. First, in Sec.~\ref{ssec_trans}, we analyze the transverse 
nucleon motion by integrating over the longitudinal momenta, considering projections on both the $x$-axis,
\begin{equation}
\frac{dN}{dp_{x}}(t)=d\cdot V_{cell} \int{\frac{dp_y dp_z}{(2\pi)^3}f(E;\mu_N,T)} \ ,
\label{eq_xmom}
\end{equation}
and on the $y$-axis, 
\begin{equation}
\frac{dN}{dp_{y}}(t)=d\cdot V_{cell} \int{\frac{dp_x dp_z}{(2\pi)^3}f(E;\mu_N,T)} \ ,
\label{eq_ymom}
\end{equation} 
which allows us to check the symmetry expected for a central collision. Here, $V_{\rm cell}$=8\,fm$^3$ is the cell volume, and $d=4$ is the 
(spin-isospin) degeneracy. It turns out that a thermal ansatz for the fit function $f$ yields fair agreement with the transport data for the 
transverse-momentum distributions, leaving 2 free parameters in terms of the chemical potential, $\mu_N$, and temperature, $T$. 
However, we enforce exact number conservation in each cell, which reduces the fit to a single parameter (chosen as the temperature).

The longitudinal momentum distributions turn out to have a more complex structure as a result of the initial beam momentum and the two nuclei entering 
the central cells at different times depending on their location. To account for the effects of two kinematically independent nuclei colliding, we employ a 
two-centroid distribution around momenta $p_{01,02}$ with weight parameter $w$ written as
\begin{multline}
\frac{dN}{dp_z}(t)=d\cdot V_{cell}\\
\bigg(\int\frac{\sqrt{\xi_1}d^2p_\perp}{(2\pi)^3}\frac{w}{1+\exp[\frac{p_\perp^2+\xi_1(p_z-p_{01})^2-2m\mu_N}{2mT}]}\\
+\int{\frac{\sqrt{\xi_2}d^2p_\perp}{(2\pi)^3}\frac{1-w}{1+\exp[\frac{p_\perp^2+\xi_2(p_z-p_{02})^2-2m\mu_N}{2mT}]}}\bigg) \ .
\label{eq_longmom}
\end{multline}
This fit function utilizes the extracted thermal parameters ($\mu_N$, $T$) from the transverse direction to allow comparisons between 
longitudinal and transverse dynamics. The purpose of the centroid momenta is to quantify how much momentum from the collective motion of 
the nucleons as part of incoming nuclei is still present after the nuclei start interacting, where the weight parameter, $w$, characterizes the number 
of nucleons from each centroid in the cell. In addition, we introduce thermal ``stretch" parameters, $\xi_{1,2}$, for the longitudinal direction of each 
centroid to account for a thermal ``anisotropy" between the transverse and longitudinal directions.  This will turn out to be required to account for the 
narrower momentum distributions in the $z$ direction, relative to the transverse ones, as observed in the transport simulations, while keeping their 
normalization conserved through the $\sqrt{\xi}$ factors in the numerator. We note that this entails a violation of the Pauli principle as the distributions 
can acquire values larger than one. However, when instead dropping these factors, we would need to increase the chemical potential to preserve the
correct density, which in turn increases the width in the transverse distributions and thus would prevent us from obtaining simultaneous fits to the 
longitudinal and transverse distributions. A violation of the Pauli principle at a level of up to $\sim$20\% has been noted in the transport simulations
early on~\cite{Papa:2000ef}. The stronger effects found in our fit functions might be caused by our relatively simple ansatz of a non-interacting Fermi 
gas, and by neglecting the spatially dependent Coulomb mean fields present in the simulations.

\subsection{Transverse Momentum Kinetics}
\label{ssec_trans}

\begin{figure}[thb!]
\begin{minipage}{1.0\linewidth}
\centering
\includegraphics[width=\linewidth]{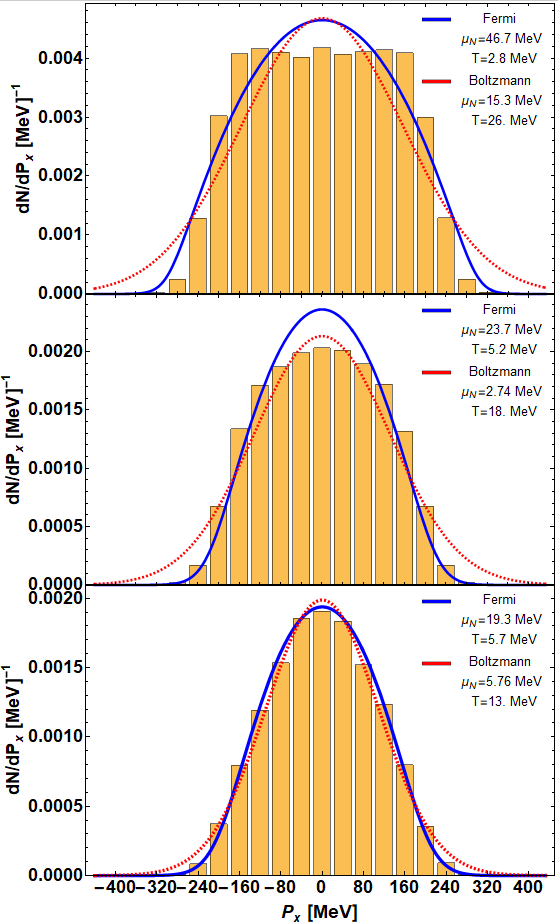} 
\end{minipage}
\caption{Results of our fits using Fermi (blue solid lines) or Boltzmann (red dotted lines) distributions for the transverse-momentum distributions in the $x$-direction, compared to the CoMD transport outputs (orange histograms) for time snapshots at 65\,fm/$c$ (top panel, corresponding to maximal density), 115\,fm/$c$ (middle panel, near the start of the temperature plateau) and 175\,fm/$c$ (bottom panel, near the onset of 3D isotropy).} 
\label{fig_comparison}
\end{figure}

\begin{figure}[thb!]
\begin{minipage}{1.0\linewidth}
\centering
\includegraphics[width=\linewidth]{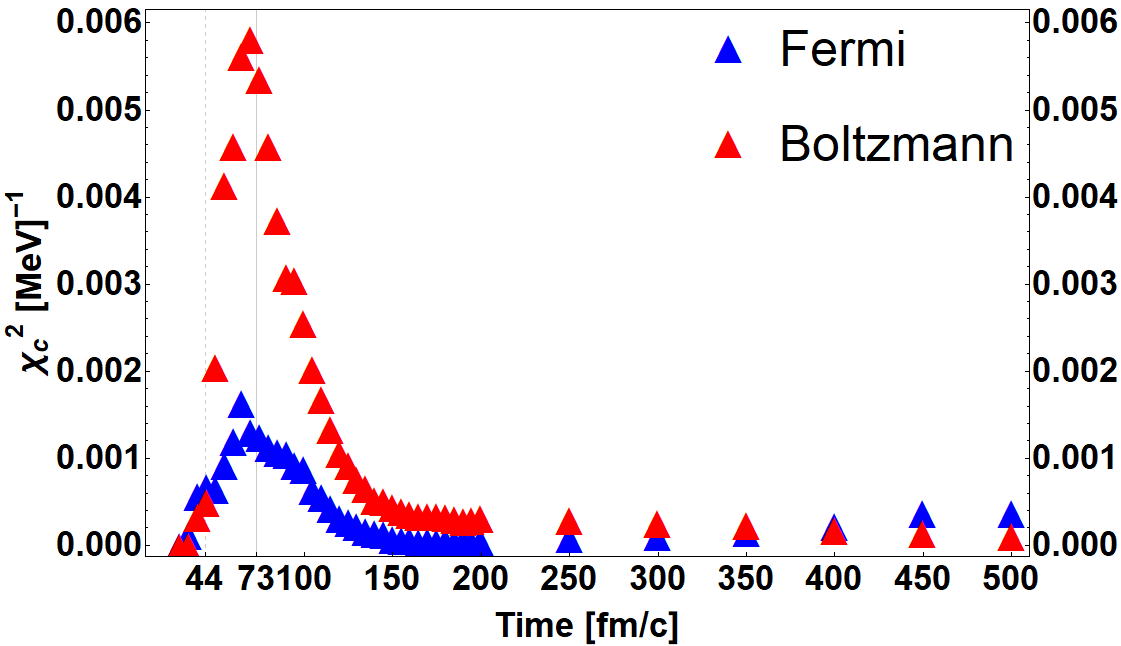} 
\end{minipage}
\caption{Time evolution of the fit quality characterized by $\chi_c^2$ for Fermi (blue) or Boltzmann statistics (red).} 
\label{fig_chisquare}
\end{figure}
Each momentum distribution in the $x$-direction is tabulated at each time step of 5\,fm/$c$, compiled by averaging over data from all 8 cells in the available octants (taking advantage of their symmetry properties). To begin with, we have utilized both thermal Boltzmann and Fermi distributions to probe the sensitivity to effects of quantum 
statistics. In Fig.~\ref{fig_comparison} we display a selection of comparisons between fit functions and histograms from the transport output for both distributions, for 3 snapshots of time roughly representing the maximal compression phase, the onset of a density plateau, and near the 3D equilibration (as will be seen later). The specific fitting procedure minimizes the relative difference, $R$, between the fit function and the values of the histogram, 
\begin{equation}
    R(t)=\sum_i\frac{\lvert\text{fit}(t,i)-\text{histogram}(t,i)\rvert}{\text{histogram}(t,i)} \ .
\label{eq_relerror}
\end{equation}
This procedure yields a more accurate representation of the number of nucleons in the high-momentum tails of the spectra (which is important for future applications such as the calculation of photon spectra). The accuracy of the fit functions is quantified using the $\chi_c^2$ value of the fit functions at each timestep,
\begin{equation}
    \chi_c^2(t)=\sum_i\frac{\big(\text{fit}(t,i)-\text{histogram}(t,i)\big)^2}{\text{histogram}(t,i)} \ ,
\label{eq_chisquare}
\end{equation}
and is shown in Fig.~\ref{fig_chisquare}. These results confirm the visual impression that fits based on Fermi distributions are superior to those 
with Boltzmann ones. This is particularly pronounced in the earlier stages of the collision, where the compression is largest but the temperatures 
are still relatively low, \ie, under conditions where one expects the effects of Pauli blocking to be most relevant.

\begin{figure}[hbt!]
\begin{minipage}{1.0\linewidth}
\centering
\includegraphics[width=\linewidth]{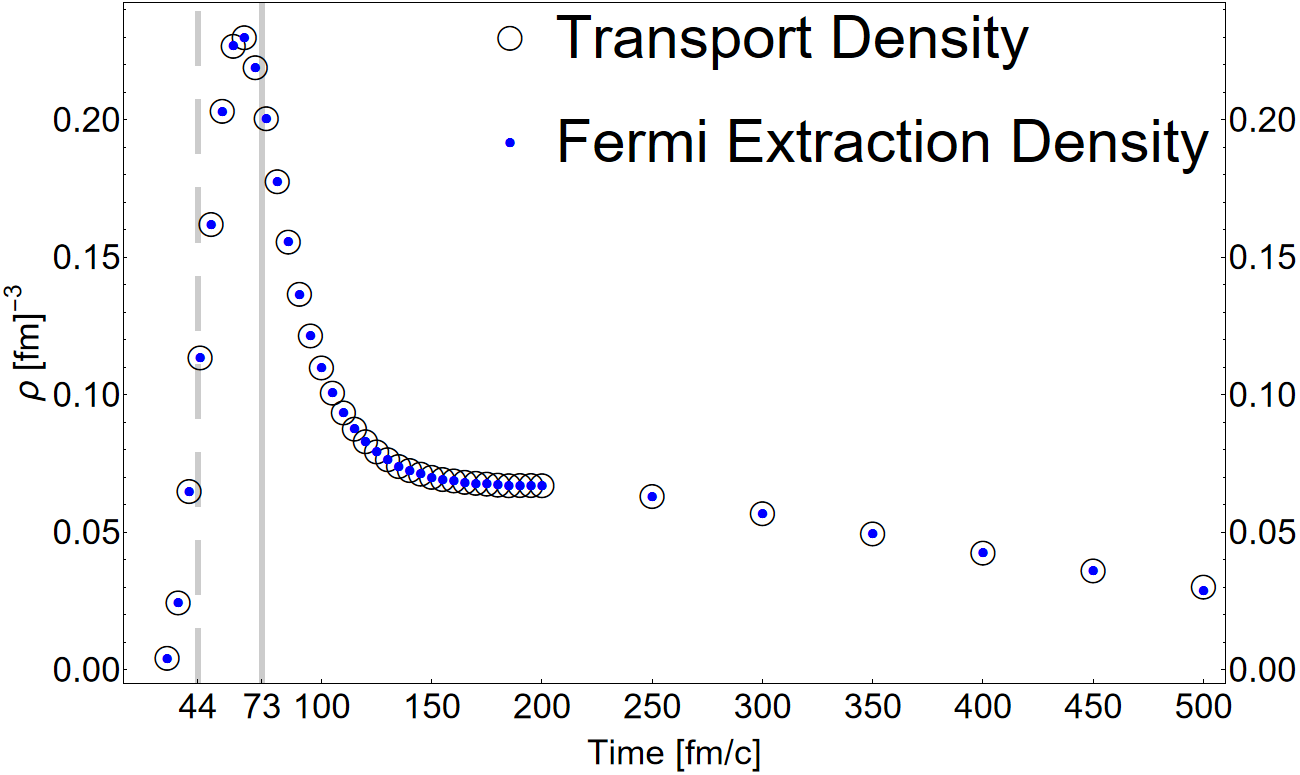} 
\end{minipage}
\caption{Comparison between the time evolution of nucleon density calculated from the transport output (black circles) and from integrating the Fermi fit functions with time-dependent extracted parameters (blue dots). The dashed line marks the approximate time when the nuclei would touch when treated as classical spheres, and the solid line indicates when the two nuclei would reach full geometric overlap if there were no interactions.} 
\label{fig_density}
\end{figure}

To check our implementation of particle number conservation, we compare in Fig.~\ref{fig_density} the nucleon density following from the fit 
parameters, 
\begin{equation}
\rho=d\int{\frac{d^3p}{(2\pi)^3}f(E;\mu_N,T)} \ ,
\label{eq_density}
\end{equation}
with the one from the transport output, where we counted the number of nucleons whose positions were in the cell and then divided by the volume of the cell, referred to as the ``transport density." The agreement is satisfactory. The graph also shows the density to reach a maximum of near 1.5 times normal nuclear matter density ($\rho_0=0.16$\,fm$^{-3}$) at about 20~fm/$c$ after initial impact, a leveling off after about 100~fm/$c$, followed by a further drop after about 200\,fm/$c$ (note that the initial impact in our definition of evolution time used in the plots occurs at about 44~fm/$c$).

\begin{figure}[bht!]
\begin{minipage}{1.0\linewidth}
\centering
\includegraphics[width=\linewidth]{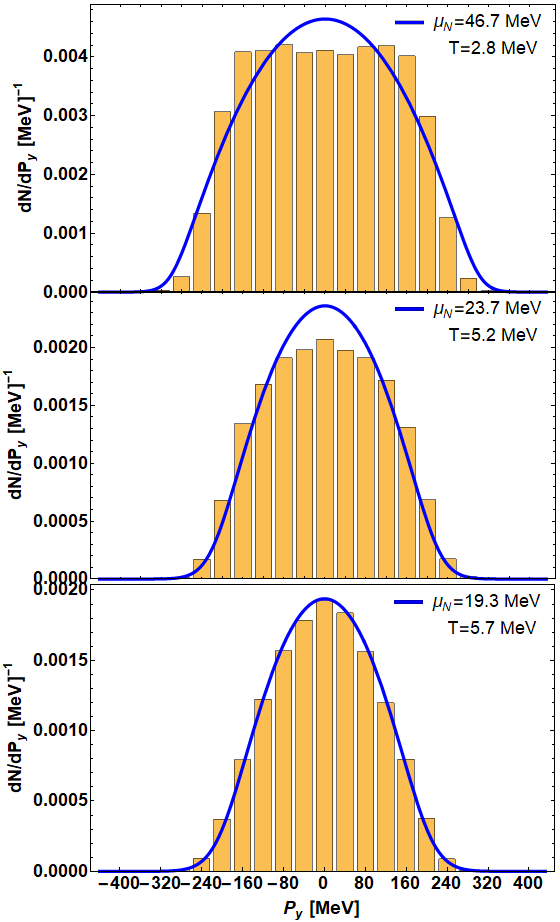} 
\end{minipage}
\caption{Comparison of our Fermi fit functions using parameters from the $x$-direction to the transport distributions in the $y$-direction for time snapshots 
at 65\,fm/$c$ (top panel), 115\,fm/$c$ (middle panel), and 175\,fm/$c$ (bottom panel).} 
\label{fig_ymom}
\end{figure}

Based on the extracted fit parameters from the momentum distributions in the $x$-direction, we show in Fig.~\ref{fig_ymom} snapshots of the momentum distributions in the $y$-direction. Good agreement is found supporting the azimuthal symmetry assumed in our extraction for central collisions.

\subsection{Longitudinal Momentum Kinetics}
\label{ssec_long}
\begin{figure}[htb!]
\begin{minipage}{1.0\linewidth}
\centering
\includegraphics[width=\linewidth]{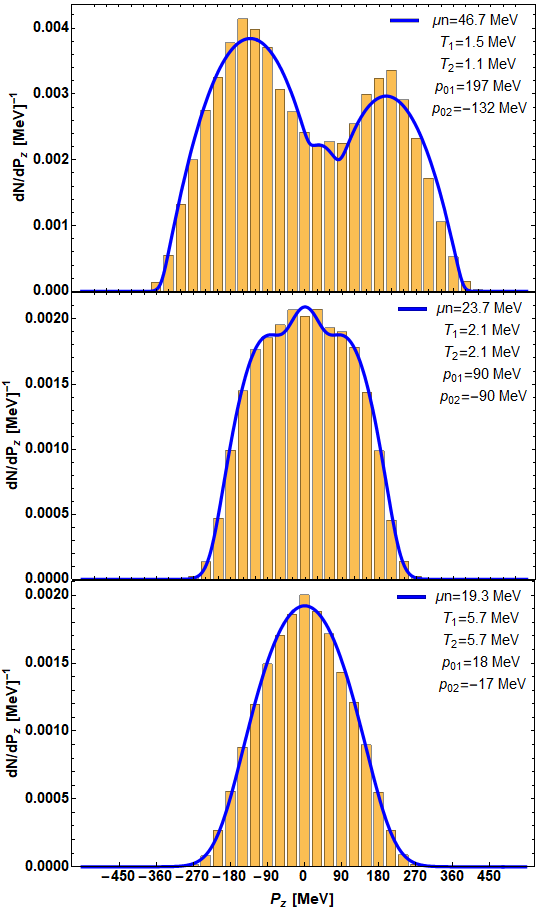}
\end{minipage}
\caption{Comparison of our Fermi fit functions for the longitudinal momentum distributions (blues lines) to the transport distributions (orange histograms) for time snapshots at 65\,fm/$c$ (top panel; maximal compression), 115\,fm/$c$ (middle panel, onset of temperature plateau), and 175\,fm/$c$ (bottom panel; onset of isotropy).}
\label{fig_longmom}
\end{figure}

\begin{figure}[bht!]
\begin{minipage}{1.0\linewidth}
\centering
\includegraphics[width=\linewidth]{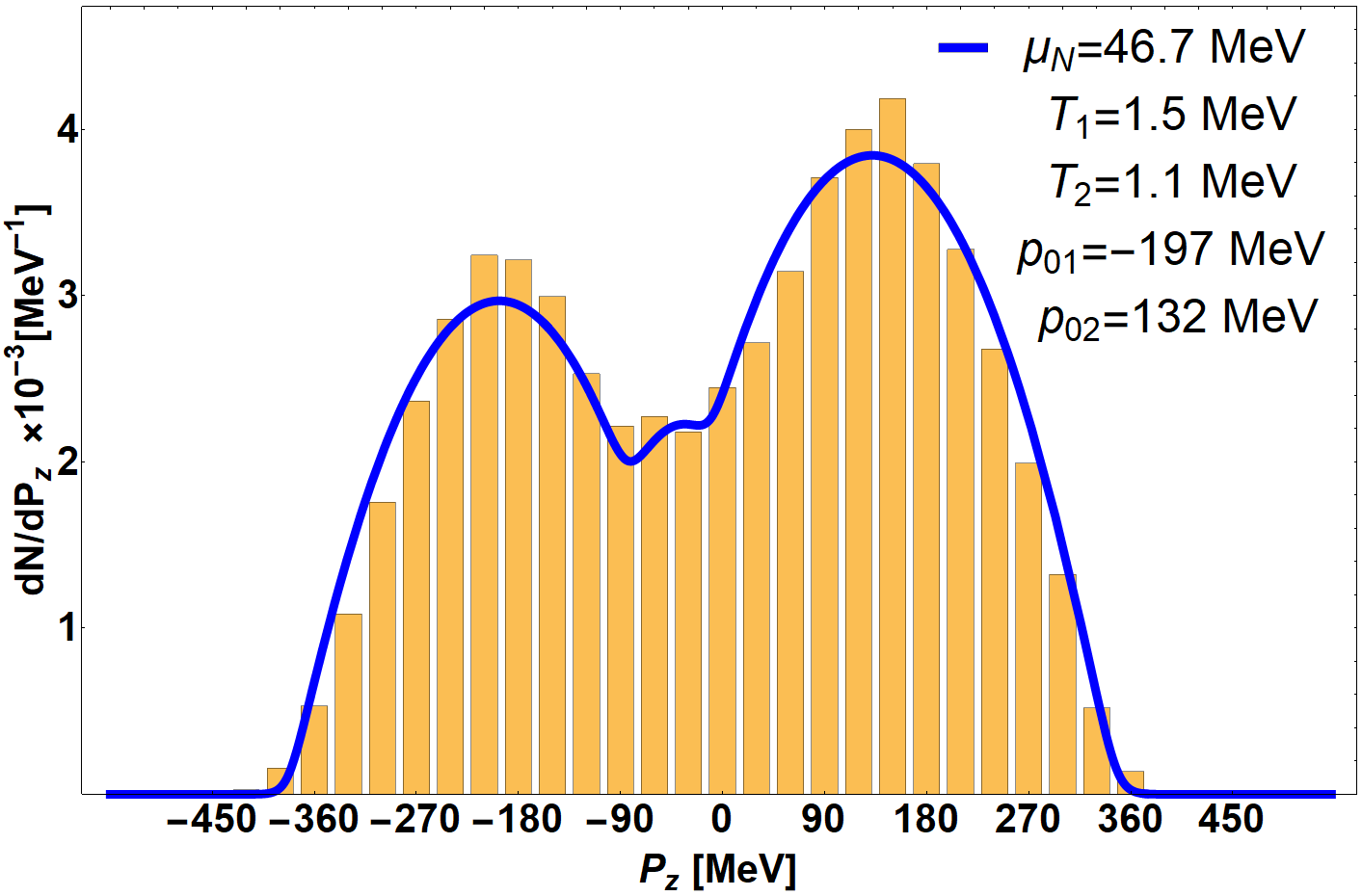} 
\end{minipage}
\caption{Comparison of our Fermi fit function for the longitudinal momentum distributions (blues lines) to the transport distributions (orange histograms) from the cells in the negative-$z$ half-space at 65\,fm/$c$ (maximal compression; $\sim$20\,fm/$c$ after nuclei touch).} 
\label{fig_mirrormom}
\end{figure}

We now turn to the longitudinal momentum distributions, along the beam direction. For early times of the collision the transport distributions have a pronounced two-hump structure that originates from the initial motion of the nuclei entering the cell under consideration, cf.~the orange histogram in the upper panel of Fig.~\ref{fig_longmom}. Clearly, this cannot be captured by a thermal fit function as was used for the transverse direction. This is the reason for introducing the two-centroid motion into the longitudinal fit function, Eq.~(\ref{eq_longmom}). This function can describe the initial two-hump structure while also transitioning into the late-time momentum distributions that have merged into a single peak, cf. lower 2 panels of Fig.~\ref{fig_longmom}. 
The difference in height between the two peaks in the upper panel of Fig.~\ref{fig_longmom}, which is for the positive-$z$ half-space, is due to the different number of nucleons from the two incoming nuclei present in a given cell at early times. We recall from  Figs.~\ref{fig_schematic} and \ref{fig_velocity} that the asymmetry between the longitudinal momentum distributions of the cells in the positive- and negative-$z$ half-spaces is a consequence of the left nucleus (the nucleus with a positive velocity) entering the cells in the negative-$z$ half-space sooner and the right nucleus (the nucleus with a negative velocity) entering the cells in the positive-$z$ half-space sooner. Therefore, the hump centered around negative $p_z$ is taller because more nucleons from the right nucleus have arrived in that cell.  At the same time the magnitude of negative centroid momentum is smaller than that of the positive one, in a way that the total momentum is zero (by construction), which is incorporated in the fits through the weight parameter $w$. The distributions in the negative-$z$ half-space have the same shape except that the hump centered around the centroid with a positive value is taller, cf.~Fig.~\ref{fig_mirrormom}. We are able to take advantage of the similarity between the shapes of the distributions of the cells in the positive- and negative-$z$ half-spaces by reflecting and then adding the distributions. On the other hand, if one adds the distributions without reflection, the resulting two-hump distribution becomes symmetric, which, however, erases kinematic information about the entrance channels of the nuclei (in particular also in the transverse directions). The overall fit quality of the longitudinal momentum distributions using our off-equilibrium ansatz is fair. 

\begin{figure}[bht!]
\begin{minipage}{\linewidth}
\centering
\includegraphics[width=\linewidth]{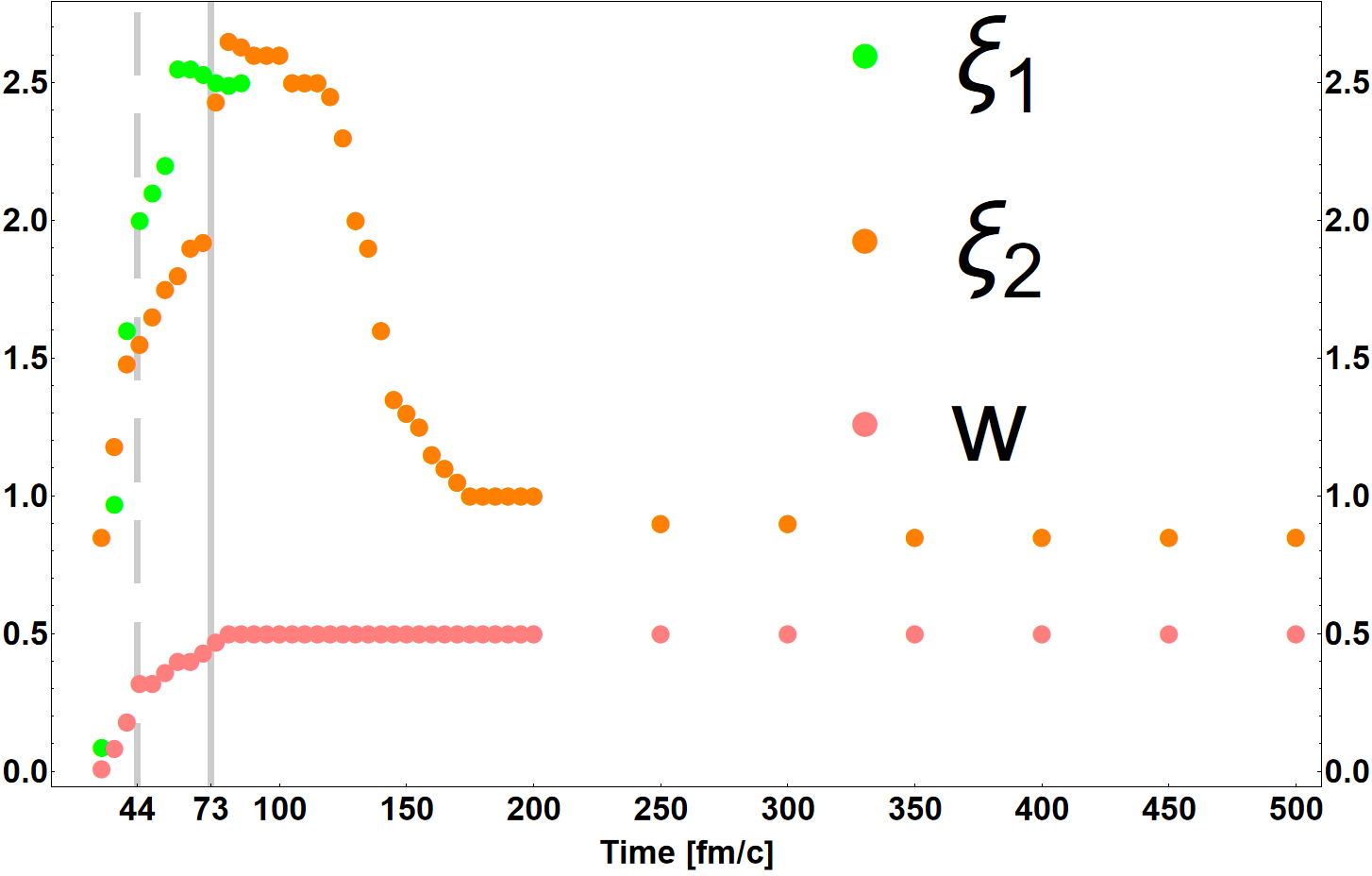}
\end{minipage}
\caption{Time evolution of the weight and thermal stretch parameters for the longitudinal temperature of the two momentum centroids from the incoming nuclei. The dashed and solid lines are the same as in Fig~\ref{fig_density}.}
\label{fig_stretch}
\end{figure}
We note that the thermal stretch parameters in Eq.~(\ref{eq_longmom}) have values greater than 1 until $t\sim$140\,fm/$c$ after the initial nuclear contact, see Fig.~\ref{fig_stretch}. This means that the effective longitudinal temperature $T_z^{\ eff} =\frac{T}{\xi}$ (as quoted in the panels of Fig.~\ref{fig_longmom}) is significantly less than the temperature extracted from the transverse direction, dictated by a narrower width of the centroid peaks, \ie, the thermal motion within the peaks is less pronounced than in the transverse direction. However, at late times the $\xi$'s reach one and approximately stay there.

\section{System Kinetics}
\label{sec_evo}
\begin{figure}[htb]
\begin{minipage}{1.0\linewidth}
\centering
\includegraphics[width=1.0\textwidth]{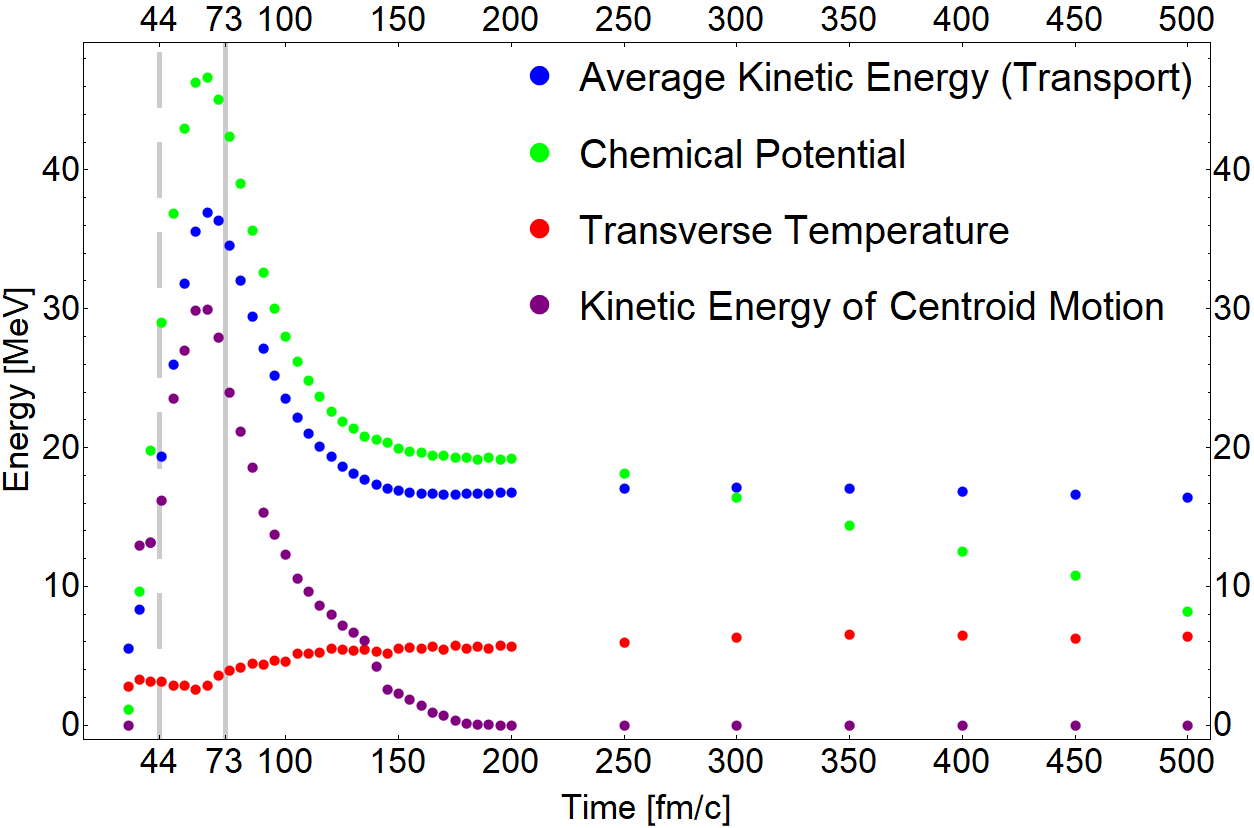}
\end{minipage}
\caption{Time evolution of thermodynamic properties of nuclear matter extracted from the central cells of CoMD transport simulations of central 35\,A$\cdot$MeV \ce{^{40}Ca} +\ce{^{40}Ca} collisions, \ie, the average total kinetic energy (blue dots), nucleon chemical potential (green dots) and temperature (red dots) from fits of the transverse-momentum spectra, and the kinetic energy of the motion of the two centroids in the longitudinal direction (purple dots). The vertical dashed and solid lines are the same as in Fig.~\ref{fig_density}.}
\label{fig_time-evo}
\end{figure}

In Fig.~\ref{fig_time-evo} we plot the time evolution of the parameters extracted from our fit functions representing the momentum distributions 
of the nucleons in cells adjacent to the center of the collision. 
We first note the rapid increase of the chemical potential, the average kinetic energy and the kinetic energy of the centroid motion (due to nuclei entering the cells). Very early on, the latter two are very close to each other, but then start to diverge signifying the redistribution of the incoming energy into thermal motion. This redistribution first occurs primarily into the transverse direction (recall that at, say, $t=115$\,fm/$c$, the transverse temperature has already heated up to about 5\,MeV while the longitudinal one is still at about only $\sim$2\,MeV), but eventually the longitudinal ``thermalization" catches up, and by the time the centroid motion has dissipated at $\sim$175\,fm/$c$, both temperatures are essentially equal, at close to 6\,MeV. Thereafter, the average kinetic energy stays approximately constant, and the medium in the central cells continues to dilute further after about $t\simeq200$\,fm/$c$ as indicated by a further drop off of the chemical potential (recall also the evolution of the number density displayed in Fig.~\ref{fig_density}). It is quite remarkable how the system establishes near local equilibrium in the transverse motion rather rapidly (not unlike what happens in ultrarelativistic heavy-ion collisions), but then also reaches 3D equilibration but on a much longer timescale (this is different from the high-energy case where the longitudinal energy does not fully dissipate, even though a large fraction of it does; also note that the situation here does not correspond to a Bjorken-type expansion where the space- and momentum-rapidities are closely correlated, since the opposite centroid momenta show up in the same cell whose size roughly corresponds to the range of the nucleon-nucleon interaction).
\begin{figure}[thb!]
\begin{minipage}{1.0\linewidth}
\centering
\includegraphics[width=\linewidth]{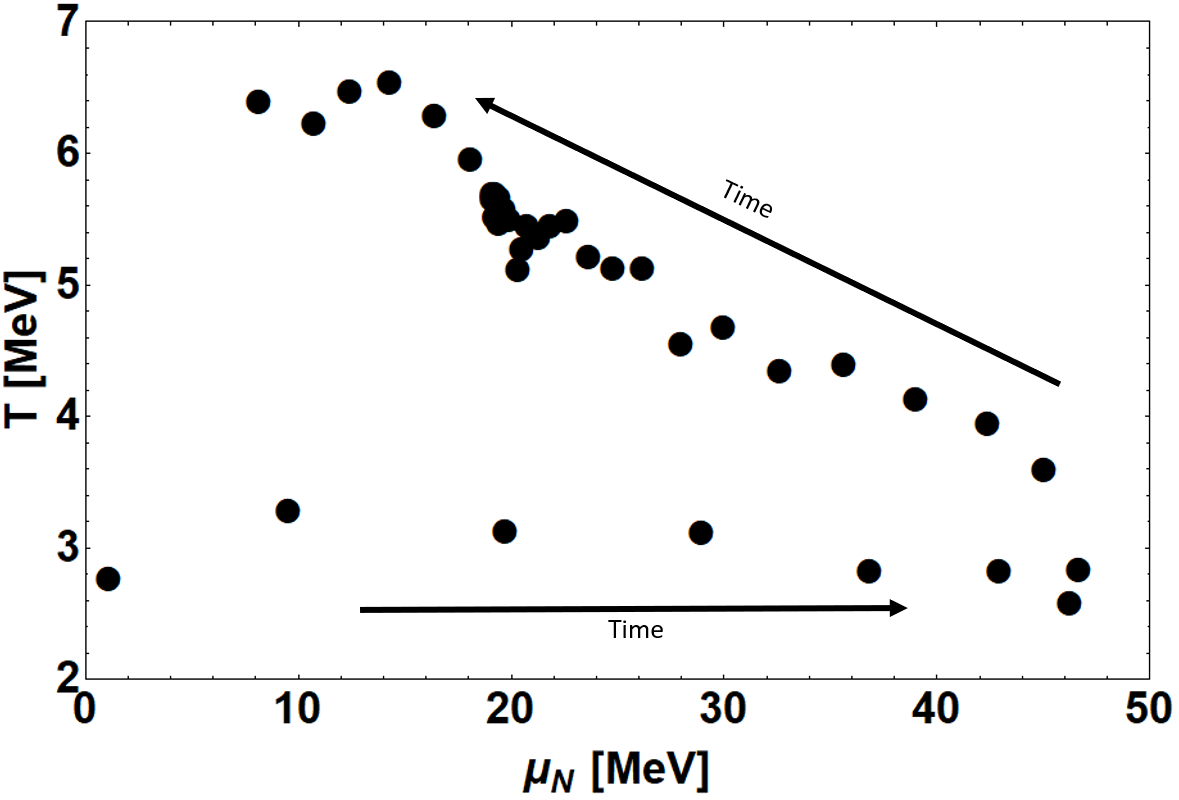}
\end{minipage}
\caption{Our results for the trajectory of the central cells in collisions of 35\,A$\cdot$MeV \ce{^{40}Ca}+\ce{^{40}Ca} in the phase diagram of strong-interaction matter.}
\label{fig_phasedia}
\end{figure}

Let us compare our results with investigations of collisions at similar energies from the literature. In Ref.~\cite{Zhou:2012bd} the time evolution of number and kinetic-energy densities, as well as temperature have been extracted from IQMD simulations of Au-Au collisions down to collision energies of 70\,MeV, by extracting bulk properties (number, kinetic-energy and entropy density) within a central sphere of radius 5\,fm (corresponding to a volume of 524\,fm$^3$). While this is significantly larger than our cells (with individual volume of 8\,fm$^3$), the main difference is that the analysis of Ref.~\cite{Zhou:2012bd} does not disentangle the effect of the incoming directed-motion energy. As a result the temperature peaks essentially synchronously with the number density, which is much earlier than in our extraction. The reason is that in our approach the incoming directed motion is eliminated from the temperature extraction through a microscopic analysis of the one-body kinematics in rest frame of each cell. For example, at the time of maximal compression, the kinetic energy of the centroids' motion still accounts for about 80\% of the total kinetic energy, and thus the temperature in the transverse direction is substantially smaller than the maximal value which is only reached much later in the evolution. The asymptotic temperatures appear to be quite similar in both calculations. In Ref.~\cite{Liu:2017vlx} thermal and transport parameters were analyzed in $^{129}$Xe + $^{119}$Sn collisions also based on IQMD but with the extra inclusion of momentum-dependent mean fields, and for a central cube of $6^3$\,fm$^3$; specifically, the temperature was extracted from the variance of quadrupole moments of the nucleon momentum distributions in the transverse plane only. Also here the temperature reaches its maximum only marginally later (ca. 10\,fm/$c$) than the number density, at a value of $\sim$6\,MeV for $E_{\rm lab}$=35\,A$\cdot$MeV. This is very comparable to the maximum transverse temperatures found in the present work, although the latter are reached at much later times. On the other hand, the large-time values of the temperature in Ref.~\cite{Liu:2017vlx} are significantly smaller than in the present work, potentially due to the momentum-dependent potentials.

Finally we represent our results as a trajectory in the phase diagram, \ie, in the temperature vs.~chemical-potential plane, cf.~Fig.~\ref{fig_phasedia}. 
The trajectory directly reflects the 4 main stages of the fireball evolution mentioned before: an initial compression at small and near-constant temperature, an initial decompression associated with significant heating from $T\simeq3$\,MeV to $\sim$5.5\,MeV, a duration of approximately constant $T$ and $\mu_N$, and a final dilution with a slight additional heating. The overall heating trend, associated with entropy production, is quite different from the trajectories typically encountered in collisions at ultra-/relativistic collision energies. 

\section{Conclusion}
\label{sec_concl}
In this paper we have introduced a coarse-graining method for heavy-ion collisions at Fermi energies where thermally-motivated distribution functions were used to fit the single-nucleon distribution functions obtained from CoMD transport simulations. While the transverse-momentum distributions turned out to be amenable to fits with Fermi distribution functions, the longitudinal distributions exhibit a strong imprint of the initial motion of the incoming nuclei. By employing a non-equilibrium ansatz for the $p_z$ dependence of the distribution functions to account for the time-dependent motion of the pertinent momentum centroids, we were able to achieve a reasonably accurate description of the 3-dimensional distributions obtained from the transport simulations over the entire evolution of the fireball. The introduction of the centroid motion, together with an extraction of the thermal parameters in the rest frame of each cell, plays a critical role in characterizing the thermo-kinetic properties of the system. In particular, it enabled us to systematically track the conversion of the incoming longitudinal energy into thermal motion. As a result, we have found significantly smaller temperatures in the early phases of the fireball compared to previous studies, building up gradually to about 6\,MeV in the late states of central 35\,A$\cdot$MeV $^{40}$Ca+$^{40}$Ca collisions. We have also found that at the maximal compression of the system the transverse temperature has reached less than half of its maximal value and even less for the longitudinal direction. Yet, after about 150\,fm/$c$, the initial longitudinal (directed) energy of the incoming nuclei has dissipated, and the system has both isotropized and thermalized in the local rest frame. Going forward, we plan to utilize our results to investigate the production of photons in heavy-ion collisions at Fermi energies, by constructing thermal production rates under the inclusion of the non-equilibrium effects introduced here, and convoluting them over the time evolution of the collision. In this way we can complement existing methods which are mostly based on probabilistic implementations of individual nucleon-nucleon collisions in 
transport simulations.

\acknowledgments
This work is supported by the Department of Energy and National Nuclear Security Administration under grant no. DE-NA 0003841 (C.E.N.T.A.U.R.)
and grant no. DE-FG03-93ER40773.

\bibliography{references}

\end{document}